\documentstyle[emulate_apj,apjfonts,epsf]{article}

\input epsf.sty
\newcommand{\mincir}{\raise -2.truept\hbox{\rlap{\hbox{$\sim$}}\raise5.truept
\hbox{$<$}\ }}
\newcommand{\magcir}{\raise -2.truept\hbox{\rlap{\hbox{$\sim$}}\raise5.truept
\hbox{$>$}\ }}
\newcommand{\siml}{\raise -2.truept\hbox{\rlap{\hbox{$\sim$}}\raise5.truept
\hbox{$<$}\ }}
\newcommand{\simg}{\raise -2.truept\hbox{\rlap{\hbox{$\sim$}}\raise5.truept
\hbox{$>$}\ }}

\newcommand{\be}{\begin{equation}}
\newcommand{\ee}{\end{equation}}
\newcommand{\ba}{\begin{eqnarray}}
\newcommand{\ea}{\end{eqnarray}}
\newcommand{\brr}{\begin{array}}
\newcommand{\err}{\end{array}}
\newcommand{\bc}{\begin{center}}
\newcommand{\ec}{\end{center}}
\newcommand{\lb}{{\left<\right.}}
\newcommand{\rb}{{\left.\right>}}

\lefthead{Borgani et al.}
\righthead{Pre--heating the ICM}
\slugcomment{To appear in The Astrophysical Journal Letters}
\begin{document}

\title{Pre-heating the ICM in high resolution simulations:
the effect on the gas entropy}

\author{S. Borgani\altaffilmark{1}, F. Governato\altaffilmark{2}, 
J. Wadsley\altaffilmark{3}, N. Menci\altaffilmark{4}, P. Tozzi\altaffilmark{5}, G. Lake\altaffilmark{6}, T. Quinn\altaffilmark{6} \&
J. Stadel\altaffilmark{6}}

\altaffiltext{1} {INFN, Sezione di Trieste, c/o Dipartimento di Astronomia
  dell'Universit\`a, via G. Tiepolo 11, I-34131 Trieste, Italy
(borgani@ts.astro.it)}
\altaffiltext{2} {Osservatorio Astronomico di Brera, via Brera 28, I-20131,
Milano, Italy (fabio@merate.mi.astro.it)}
\altaffiltext{3} {Department of Physics and Astronomy, McMaster University,
Hamilton, Ontario, L88 4M1, Canada (wadsley@physics.mcmaster.ca)}
\altaffiltext{4} {Osservatorio Astronomico di Roma, via
dell'Osservatorio, I-00040 Monteporzio, Italy (menci@coma.mporzio.astro.it)}
\altaffiltext{5} {Osservatorio Astronomico di Trieste, via Tiepolo 11, I-34131
Trieste, Italy (tozzi@ts.astro.it)}
\altaffiltext{6} {Department of Astronomy, University of Washington, Seattle WA
98195, USA (lake, trq, stadel@astro.washington.edu)}

\begin{abstract}
We present first results from high--resolution Tree+SPH simulations of
galaxy clusters and groups, aimed at studying the effect of
non--gravitational heating on the entropy of the intra--cluster medium
(ICM). We simulate three systems, having emission--weighted 
temperature $T_{\rm ew}\simeq$0.6,1 and 3 keV, with spatial resolution
better than 1\% of the virial radius. We consider the effect of
different prescriptions for non--gravitational ICM heating, such as
supernova (SN) energy feedback, as predicted by semi--analytical
models of galaxy formation, and two different minimum entropy floors,
$S_{\rm fl}=50$ and 100 keV cm$^{2}$, imposed at $z=3$. Simulations
with only gravitational heating nicely reproduce predictions from
self--similar ICM models, while extra heating is shown to break the
self--similarity, by a degree which depends on total injected energy
and on cluster mass. We use observational results on the excess
entropy in central regions of galaxy systems, to constrain the amount
of extra--heating required. We find that setting the entropy floor
$S_{\rm fl}=50$ keV cm$^{2}$, which corresponds to an extra heating
energy of about 1 keV per particle, is able to reproduce the observed
excess of ICM entropy.
\end{abstract}


\section{Introduction}

\begin{figure*}[th]
\vspace{0.3truecm}
\includegraphics{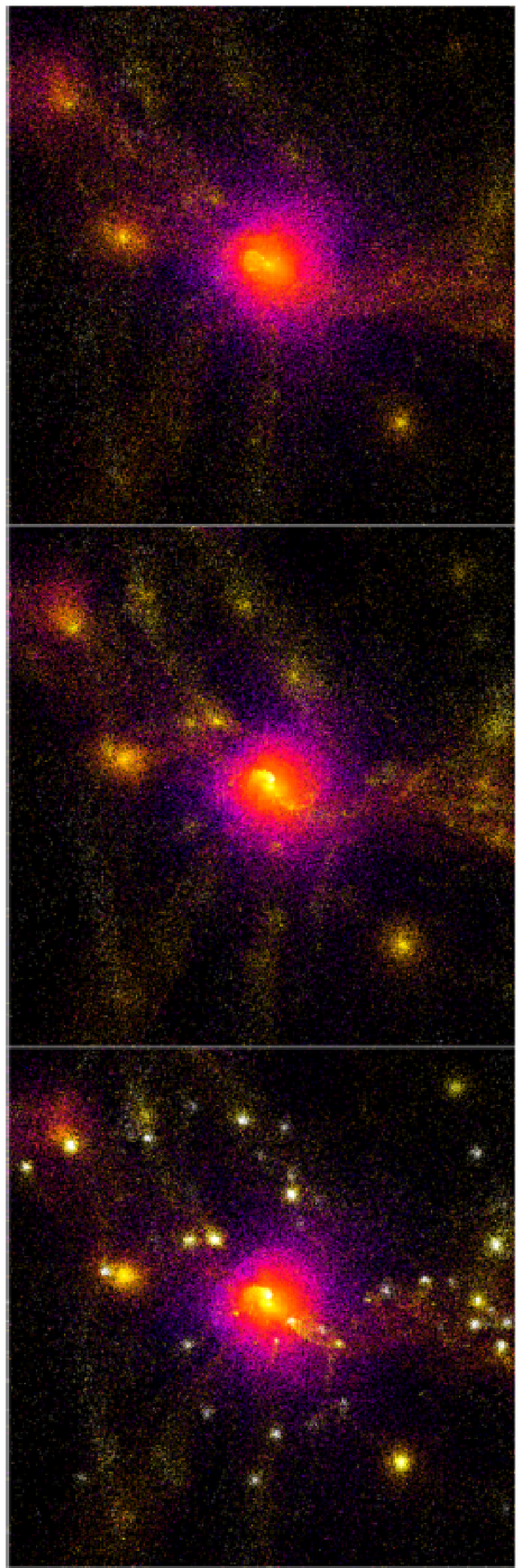}
$\ \ \ \ \ \ $\\
\vspace{5.5truecm}$\ \ \ $\\ {\small{Fig.}~1.---
Entropy maps for the Virgo cluster at $z=0$, within a
box of 12.5 Mpc. Left panel: gravitational heating only; central
panel: SN feedback with unity efficiency; right panel: entropy floor
$S_{\rm fl}=50$ keV cm$^2$ imposed at $z=3$. Brighter regions
indicate gas with lower entropy.
}
\end{figure*}

The high temperature reached by diffuse baryons within the potential
wells of galaxy clusters makes them directly observable in the
$X$--rays, mostly due to bremsstrahlung emission (e.g., Borgani \&
Guzzo 2001) With the recent advent of the Chandra--AXAF and
Newton--XMM satellites, the level of details at which the the physics
of the intra--cluster medium (ICM) can be observationally described is
undergoing an order--of--magnitude improvement, both in spatial and in
energy resolution.  From the theoretical viewpoint, the first attempt
to model the thermodynamical properties of the ICM assumed them to be
entirely determined by gravitational processes, like adiabatic
compression and shock heating (Kaiser 1986). Since gravity alone does
not introduce characteristic scales, this model predicted the gas
within clusters of different mass to behave in a self--similar
way. Under the assumption of hydrostatic equilibrium and
bremsstrahlung $X$--ray emissivity, the self--similar model predicts
the scaling $L_X\propto T^2$ between $X$--ray luminosity and gas
temperature, while $L_X\propto T^{\sim 3}$ is observed (e.g., Arnaud
\& Evrard 1999). Furthermore, defining the ICM entropy as
$S=T/n_e^{2/3}$ ($n_e$: electron number density; Ponman, Cannon \&
Navarro 1999, PCN hereafter), then self--similar scaling implies
$S\propto T$.  This is at variance with the observational evidence
that the entropy at one--tenth of the cluster virial radius tends to a
constant value, $S\sim 100$ keV cm$^2$ for $T\mincir 2$ keV, with
self--similar scaling recovered for hot, $T\magcir 6$ keV, systems
(PCN; Lloyd-Davies, Ponman \& Cannon 2000).  The current
interpretation for such discrepancies requires self--similarity to be
broken by non--gravitational gas heating (e.g., Kaiser 1991; Evrard \&
Henry 1991; Cavaliere, Menci \& Tozzi 1998; Balogh, Babul \& Patton
1999; Tozzi \& Norman 2001, TN01 hereafter; Brighenti \& Mathews
2001). This extra heating would increase the gas entropy, place it on
a higher adiabat and, therefore, prevent it from reaching high central
densities during the gravitational collapse.  Despite the general
consensus for the need of non-gravitational heating (cf. also Bryan
2000; Muanwong et al. 2001), determining the astrophysical source
responsible for it, this is still a widely debated issue. The two most
credible hypothesis are based on energy feedback from supernovae (SN;
e.g., Loewenstein 2000; Bower et al. 2001; Menci \& Cavaliere 2000) or
from AGN activity (e.g., Valageas \& Silk 1999; Wu, Fabian \& Nulsen
2000).  In this context, numerical hydrodynamical simulations
represent an invaluable tool to correctly follow dynamical
complexities, like merging of substructures and non--spherical shocks,
whose relevance for ICM properties is emphasized by recent $X$--ray
cluster observations at high spatial resolution (e.g., Markevitch et
al. 2000). Different groups have run such simulations with the aim of
understanding in details the effect of non--gravitational heating and
the amount of energy required to reproduce observations (e.g.,
Navarro, Frenk \& White 1995; Bialek, Evrard \& Mohr 2000). In
particular, Bialek et al. have run simulations at intermediate
resolution for a fairly large ensemble of clusters. After assuming
different initial values for the gas entropy, they checked the effect
of pristine extra heating on several ICM scaling relations.

In this {\em Letter} we present results from our ongoing project of
running high--resolution cluster simulations, using different schemes
for injecting non--gravitational energy feedback into the ICM. We will
concentrate here on the effect of extra--heating on the ICM entropy
and will compare the results to the observational constraints by
PCN. We reserve for a separate paper (Governato et al. 2001, Paper II
hereafter) a thorough description of the simulations and a detailed
analysis of the resulting ICM properties.

\section{The simulations}
We use GASOLINE, a parallel, multistepping tree+SPH code with periodic
boundary conditions (Wadsley, Quinn \& Stadel 2001), to re--simulate
at high resolution three halos taken from a cosmological box (100 Mpc
aside) of a $\Lambda$CDM Universe, with $\Omega_m=0.3$,
$\Omega_\Lambda=0.7$, $\sigma_8=1$, $h=H_0/(100\,{\rm km\,s}^{-1}{\rm
Mpc}^{-1})=0.7$ and $f_{\rm bar}=0.13$. In the following we give a
short descriptions of the simulations, while we refer to Paper II for
further details and for the discussion about the effect of radiative
cooling, which we neglect here. The main characteristics of the three
halos are listed in Table 1.  Owing to their virial mass and
temperature, in the following we will refer to the three simulated
structures as Virgo cluster, Fornax group and Hickson group. Thanks
to the good mass resolution, we are able to resolve with 32 particles
structures having total mass as small as about $5.5 \times
10^{10}M_\odot$ and $1.6 \times 10^{10}M_\odot$ within the Virgo
cluster and within the two smaller groups, respectively.

\vspace{6mm}
\hspace{-4mm}
\begin{minipage}{9cm}
\renewcommand{\arraystretch}{1.5}
\renewcommand{\tabcolsep}{2mm}
\begin{center}
\vspace{-3mm}
~\\ ~\\
TABLE 1\\
{\sc Characteristics of the simulated systems\\}
\footnotesize
\vspace{2mm}

\begin{tabular}{lccccc}
\hline \hline
Run & $M_{\rm vir}$ & $R_{\rm vir}$ & $T_{\rm ew}$ & $m_{gas}$ &
$\epsilon$\\ 
\hline 
Virgo   & 30.4 & 1.75 & 2.07 & 2.21 & 7.5\\ 
Fornax  & 5.91 & 1.01 & 0.95 & 0.65 & 5.0\\ 
Hickson & 2.49 & 0.76 & 0.60 & 0.65 & 5.0\\ 
\hline
\end{tabular}\\
\end{center}
\footnotesize{Column 2: total virial mass ($10^{13}M_\odot$); Column
3: virial radius (Mpc); Column 4: Emission--weighted virial
temperature (keV) for the runs including only gravitational
heating; Column 5: mass of gas particles ($10^8M_\odot$); Column 6:
Plummer--equivalent softening for gravitational force ($h^{-1}$kpc).}
\vspace{3mm}
\label{t:simul}
\end{minipage}

The first scheme for non--gravitational heating is based on setting a
minimum entropy value at some pre--collapse redshift (e.g. Navarro et
al. 1995; TN01; Bialek et al. 2000). For gas with local electron
number density $n_e$ and temperature $T$, expressed in keV, at
redshift $z$, we define the entropy as
\be 
S\,=\,{T\over n_e^{2/3}}\,=\,\left[{f_{\rm bar}\over
m_p}\,{1+X\over 2}\,\bar\rho(z)\,(1+\delta_{\rm
g})\right]^{-2/3}\,T~{\rm keV\,cm^2}\,,
\label{eq:entr}
\ee
where $\bar\rho(z)=\bar\rho_0(1+z)^3$ is the average cosmic matter
density at redshift $z$, $\delta_{\rm g}$ the gas overdensity, $m_p$
the proton mass and $X$ hydrogen mass fraction ($X=0.76$ is assumed in
the following). Accordingly, the entropy of the $i$-th gas particle in
the simulation is defined as $s_i=T_i/n_i^{2/3}$, where $T_i$ and
$n_i$ are the temperature and the electron number density associated
to that particle. At $z=3$, we select all the gas particles with
overdensity $\delta_{\rm g}>5$, so that they correspond to structures
which have already undergone turnaround. After assuming a minimum
floor entropy, $S_{\rm fl}$, each gas particle having $s_i<S_{\rm fl}$
is assigned an extra thermal energy, so as to bring its entropy to the
floor value, according to eq.(\ref{eq:entr}). We choose two values for
this entropy floor, $S_{\rm fl}=50$ and 100 keV cm$^2$.  We computed
the mean density of the heated gas at $z=3$ to be $\lb \delta_{\rm
g}\rb\simeq 185$, 280 and 215 for the Virgo, Fornax and Hickson runs,
respectively.  We assume $z_h=3$ for the reference heating redshift,
since it is close to the epoch at which sources of heating, like SN or
AGNs, are expected to reach their maximum activity. We check the
effect of changing $z_h$ by also running simulations of the Fornax
group for $z_h=1$, 2 and 5.  We estimate the amount of energy injected
in the ICM in these pre--heating schemes by selecting at $z=0$ all the
gas particles within the virial radius and tracing them back to
$z=3$. We find that taking $S_{\rm fl}=50$ keV cm$^2$ amounts to give
an average extra heating energy of $E_h={3\over 2}T_h\simeq 1.4$
keV/part for the Fornax and Hickson groups and $E_h\simeq 0.9$
keV/part for the Virgo cluster. Such values are twice as large for
$S_{\rm fl}=100$ keV cm$^2$.
We also verified that the fraction of gas particles within the virial
radius, that have been heated at $z=3$, is of about 75\%, almost
independent of the mass of the simulated system, for both values of
$S_{\rm fl}$. 

\begin{figure*}
\includegraphics{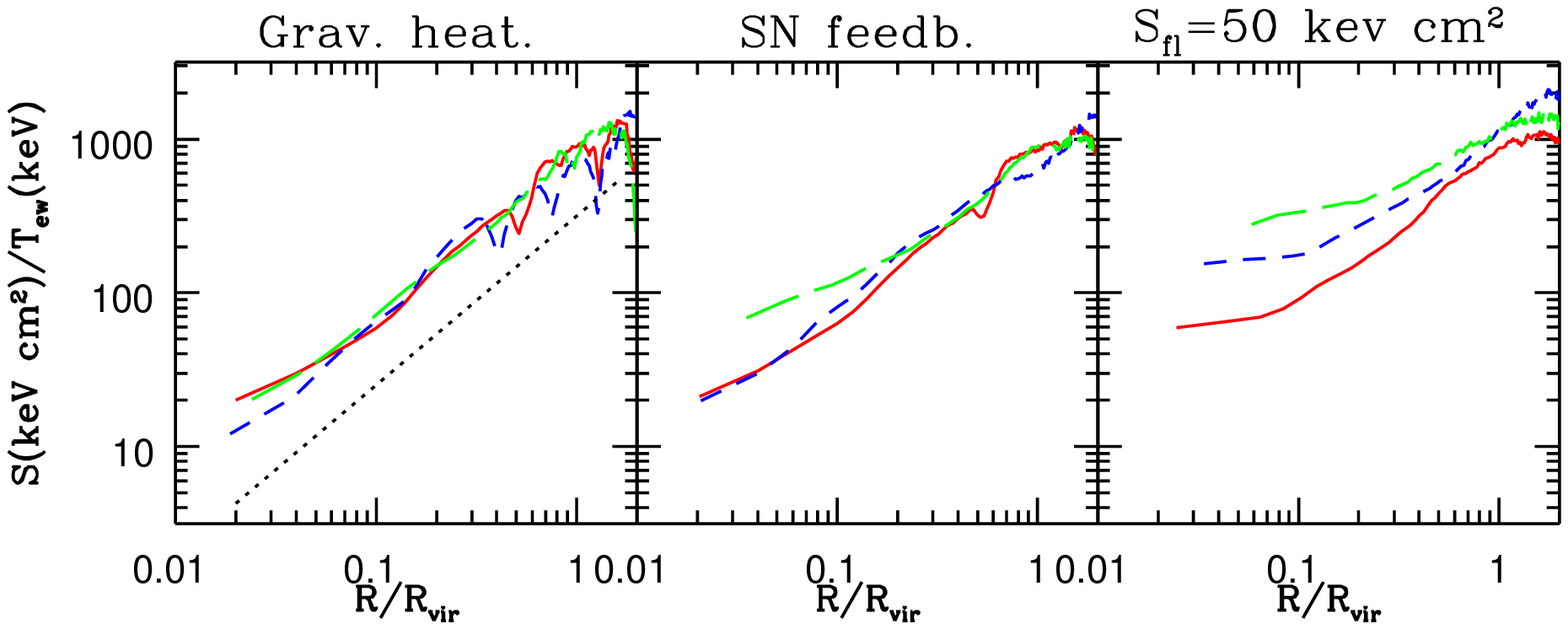}
$\ \ \ \ \ \ $\\
\vspace{5.truecm}
$\ \ \ $\\
{\small\parindent=3.5mm {Fig.}~2.--- 
Profiles of specific entropy in
units of $T_{\rm ew}$. The three panels, from left to right,
correspond to the same heating schemes as in Fig. 1. Solid,
short--dashed and long--dashed curves are for the Virgo cluster, the
Fornax group and the Hickson group, respectively. Each profile is
shown down to the radius encompassing 100 gas particles, which is
taken as the smallest scale adequately treated by the SPH scheme. The
dotted straight line in the left panel shows the analytical prediction
by TN01 for the profile of entropy induced by gravitational heating.
}
\end{figure*}

As for the pre--heating by SN feedback (e.g., Kauffmann, White \&
Guiderdoni 1993; Somerville \& Primack 1999; Cole et al. 2000), we
resort to semi--analytical modelling of galaxy formation to compute
the star--formation rate within halos having the same mass as the
simulated ones (see Poli et al. 1999, for a detailed description of
the method). We assume a feedback parameter ($\alpha_h=2$ in the model
by Poli et al.) so as to reproduce both the local B-band luminosity
function and the Tully--Fisher relation (see also Cole et
al. 2000). The resulting star--formation rates are used to derive the
history of energy release from type II SN. During the cluster
evolution, this energy is shared among all the gas particles having
$\delta_g\ge 50$ (see Paper II).  This density threshold, which
roughly corresponds to the density contrast at the virial radius, has
been chosen to guarantee that gas heating takes place inside
virialized regions.  We verified that final results do not change by
changing by a factor ten the above value of the limiting
$\delta_g$. Under the assumption that all the energy released by SN is
thermalized into the ICM, this scheme dumps a total amount of about
0.35 keV/part extra energy per particle.

\section{Results and discussion}
The entropy maps of Figure 1 show a qualitative description of the
effect of non--gravitational heating on the ICM entropy.  In the
absence of extra--heating (left panel) the high resolution achieved in
our simulation of a Virgo--like cluster reveals a wealth of
substructures in the entropy pattern. Small--size halos, which are the
first to have collapsed, are characterized by low entropy as a
consequence of the fact that they contain early accreted, and
therefore only weakly shocked, gas particles. A higher entropy level
characterizes, instead, the large--scale filaments, which are
surrounded by shells of recently accreted and strongly shocked
gas. The main structure of the cluster also shows a low--entropy core,
surrounded by regions of progressively higher entropy associated with
recently accreted gas: as a consequence of the continuous increase of
the total virialized cluster mass, the later the baryons are accreted,
the larger their infall velocity and, therefore, the stronger the
experienced shock. This process gives rise to an expanding shock
separating the inner gas, which sets in hydrostatic equilibrium, with
the external cooler and adiabatically compressed medium, the interface
occurring around the virial radius. Quite remarkably, small halos
merging into the cluster main body are able to keep their low--entropy
structure for a few crossing time scales, before their gas is
stripped. As a consequence, sharp structures arise well inside the
virial region, with entropy discontinuities and tail of gas stripped
from the merging subhalos by the effect of the ram pressure.  This
picture changes as the gas receives non--gravitational heating. As the
gas is placed on a higher adiabat, it is no longer able to accrete
inside the small--mass halos and, therefore, accretion shocks are
switched off. However, while the small--scale features are
progressively washed out as the amount of energy injection increases,
a halo of high--entropy, recently shocked gas still surrounds the
cluster main body. Although somewhat smoothed, some discontinuities in
the gas entropy are still visible even in the cluster central
regions. It is quite tempting to associate such features to those
recently observed by the Chandra satellite (e.g., Markevitch et
al. 2000; Fabian et al. 2000; Mazzotta et al. 2001). Although a close
comparison between such details of the ICM structure in simulations
and in observational data requires a more careful analysis, there is
little doubt that the increasing quality of $X$--ray data will soon
permit the reconstruction of the thermodynamic history of the
intra--cluster gas.

A more quantitative look at the ICM entropy is provided by Figure 2,
where we show the entropy profiles for the different schemes of ICM
extra--heating. By plotting the entropy, in units of
emission--weighted temperature $T_{\rm ew}$, as a function of the
radius, in units of the virial radius $R_{\rm vir}$, we emphasize the
self--similar behavior of the ICM in the presence of gravitational
heating only (the plotted quantity would be proportional to $\rho_{\rm
gas}^{-2/3}$ if the gas were isothermal).  Indeed, the profiles for
the three structures do coincide to a good accuracy in the absence of
any extra heating. In this case, the shock model developed by TN01
under the assumption of spherical accretion, predicts the entropy
profile $S\propto R^{1.1}$. This scaling is shown by the dotted line
in the left panel of Fig. 2, and nicely agrees with the scaling found
in the simulations with gravitational heating only. This agreement and
the absence of any significant flattening of the entropy profiles at
small radii are witnessing that our simulations are correctly
capturing gravitational shocks and do not produce numerical artifacts
over the considered range of scales.  The wiggles in the entropy
profiles mark the positions of the small--scale merging sub--halos,
appearing in Fig. 1, which bring low--entropy gas inside the main body
of the clusters. Although such merging structures violate the
assumption of spherical accretion of the TN01 model, they do not alter
the global behavior of the cluster entropy.  The presence of
non--gravitational heating has the twofold effect of making the
entropy profiles shallower, while breaking the self--similarity to a
degree which depends on the amount of injected extra energy. As
expected, the effect of heating is more pronounced for the halo with
the smallest virial temperature, for which the extra--energy per
particle corresponds to a larger fraction of the total virial
temperature.  As for the case with $S_{\rm fl}=50$ keV cm$^2$, the
floor value is almost recovered in the innermost resolved region only
for the Virgo simulation, while it is significantly larger for the two
smaller groups. Since the effect of the heating is that of decreasing
the gas density, a significant fraction of the shocked gas can now
flow down to the cluster central regions, thus further increasing the
entropy level.  As for the heating by SN feedback, its effect is only
marginal on the Virgo cluster and on the Fornax group, while it
significantly changes the entropy profile of the smaller Hickson group
below $\sim 0.2R_{\rm vir}$.  In general, although extra heating
largely modifies gas entropy in the cluster central regions, it has
only a marginal effect at $\sim R_{\rm vir}$. This is consistent with
the expectation that gravitational shocks provide the dominant
mechanism for establishing the global heating of the gas.

We compare in Figure 3 the observational data by PCN on the gas
entropy for clusters and groups at $0.1R_{\rm vir}$ with the results
obtained from our simulations. As discussed by PCN, such data indicates
that some pre--heating should have established an excess entropy in
central cluster regions, which causes the flattening of the $S$--$T$
relation for low--temperature systems, while being negligible for the
more massive systems, whose gas has been mainly heated by
gravitational processes.  Again, in the presence of gravitational
heating only, our simulations nicely reproduce the expectations from
self--similar scaling $S\propto T_{\rm ew}$. 
The agreement with the prediction of self--similar scaling confirms
once more that the resolution of our simulations is more than adequate
to correctly capture global ICM thermodynamical properties.  As
expected, adding extra heating breaks the self--similarity and
increases central entropy by a larger amount for smaller systems.
Heating with about one-third of keV per particle, with redshift
modulation as predicted by our SN model, has a quite small effect and
is not adequate to reproduce observational results. Both heating
recipes, based on setting an entropy floor at $z=3$, have a much
larger effect on the entropy, while still leaving $T_{\rm ew}$ almost
unchanged.  Even for the Hickson group, the value of $S(0.1R_{\rm
vir}$ for the such heating schemes turns out to be about twice as
large as $S_{\rm fl}$. This shows how after the gas is pre--heated,
gravitational effects still act to increase its entropy down to
$0.1R_{\rm vir}$. We also find that varying the heating redshift from
$z_h=1$ to $z_h=5$ in the Fornax simulation does not affect the
central entropy, thus indicating that this quantity mainly depends on
the entropy floor level and not on the epoch at which it is
established.

As a general conclusion, our results show that observational data on
the excess gas entropy in central regions of small clusters and groups
require a non--gravitational energy injection of about 1 keV per
particle. This result is in qualitative agreement with that derived by
comparing simulation results with observational data on the slope of
the luminosity--temperature relation of clusters and groups (Bialek et
al. 2000; Paper II).  Which are the implications of this result on the
astrophysical sources responsible for the heating of the
inter--galactic medium?  Although our results suggest that the
majority of the required energy budget can not be supplied by type II
SN, the final answer to the above question requires a better
understanding and a more accurate treatment of several physical
processes. Radiative cooling, which is not included in the simulations
presented here, has been advocated as a possible solution to the
problem of ICM excess entropy (e.g., Bryan 2000). Gas undergoing
cooling in central cluster regions is converted into collisionless
stars, and, as such, does not provide pressure support. Therefore,
strongly shocked gas in the cluster outskirts starts flowing inside,
thus increasing the central entropy level. Since the cooling time
scale is always much shorter than the dynamical time scale in central
cluster regions, a fraction of gas as large as $\sim 50\%$ can leave
the diffuse hot phase (e.g., Balogh et al. 2001; Paper II). This
result is at variance with observations which, instead, indicate only
a small fraction, $\sim 10\%$, of cluster baryons to reside in the
cold phase (Balogh et al. 2001), thus calling for the presence of a
feedback mechanism, which were able to prevent this ``cooling
crisis''. Furthermore, a detailed understanding of the process of the
diffusion of the energy feedback into the ICM and of the relative role
played by different heating sources are far from being reached.  In
this respect, the improvement of observational data on the abundance
and spatial distribution of heavy elements from Chandra and Newton-XMM
satellites (e.g., B\"ohringer et al. 2000) will shed light on the
interplay between ICM physics and the history of star formation in
clusters.

\includegraphics{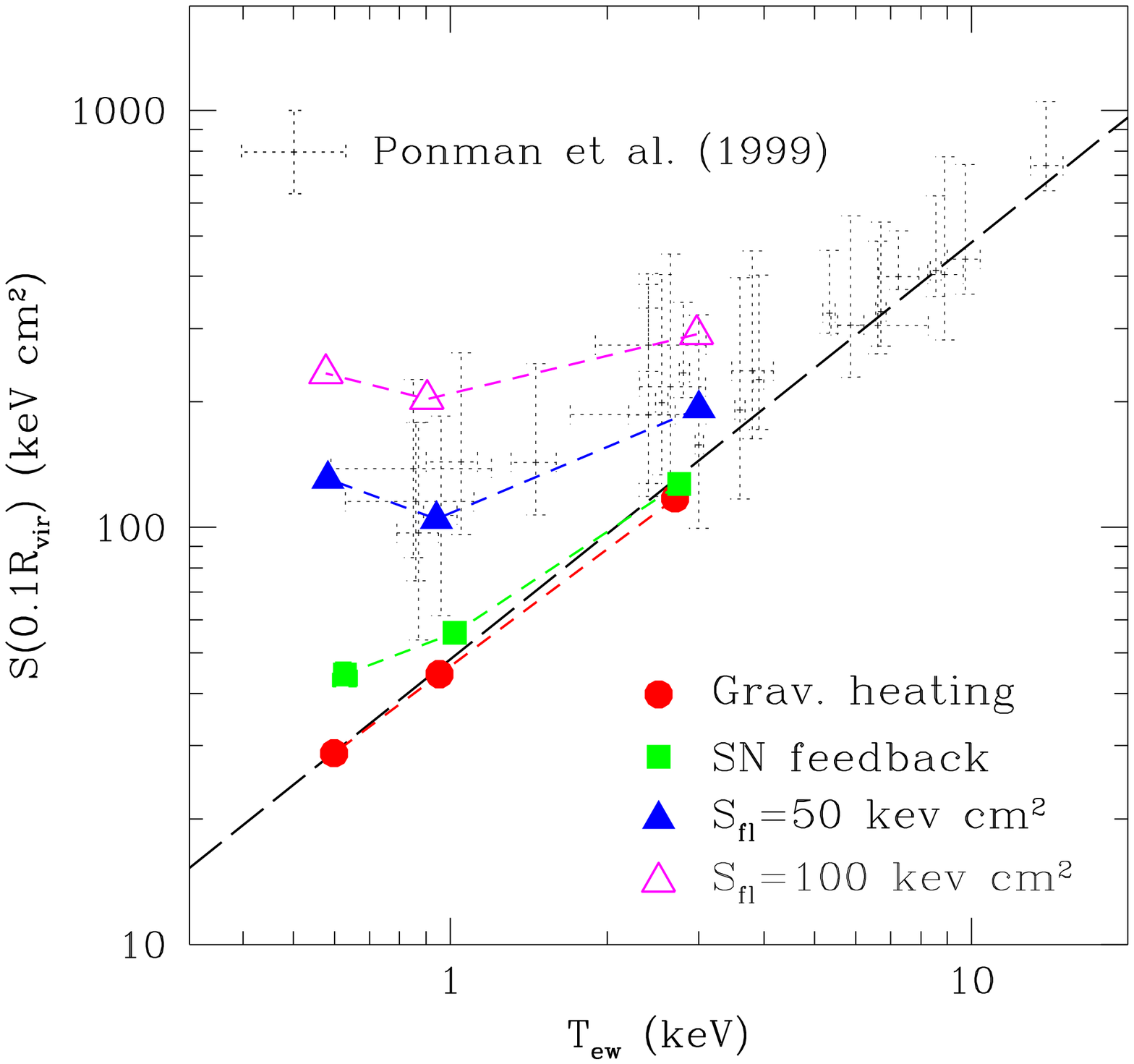} $\ \ \ \ \ \ $\\
\vspace{7.3truecm}
$\ \ \ $\\
{\small\parindent=3.5mm {Fig.}~3.--- 
The relation between specific gas entropy at
$0.1\,R_{\rm vir}$ and the emission--weighted virial temperature,
$T_{\rm ew}$. Circles are for the runs with gravitational heating,
squares for SN energy feedback, while filled and open triangles for an
entropy threshold settled at $z=3$ at the two different values
reported in the labels. Dotted crosses are the data by PCN, rescaled
to $h=0.7$. The long--dashed line shows the relation $S(0.1R_{\rm
vir})=50({T/{\rm keV})\,(f_{\rm bar}/ 0.06\,h^{-2}})^{-2/3}{\rm
keV\,cm}^2$, which fits the observational results for $T\magcir 6$ keV
clusters (Ponman et al. 1999).
}
\vspace{5mm}

\acknowledgements{The simulations have been realized at CINECA
(Bologna) and ARSC (Fairbanks) supercomputing centers. We thank
A. Cavaliere for reading the manuscript.}

\end{document}